%% file: povm_randomness_arXiv.tex
\documentclass[twocolumn,pra,aps,showpacs,superscriptaddress,floatfix,showkeys,nofootinbib,10pt]{revtex4-2}

\usepackage{amsmath,amsfonts,amssymb,amsthm}
\usepackage{graphicx}
\usepackage{tikz}
\usepackage{makecell}
\usepackage{hyperref}

\newcommand{\ket}[1]{ | \, #1 \rangle} \newcommand{\bra}[1]{ \langle #1 \, |} 
\newcommand{\proj}[1]{\ket{#1}\bra{#1}}

\newcommand{\bkc}[2]{ \langle #1 , #2 \rangle}

\newcommand{\be}{\begin{equation}} \newcommand{\ee}{\end{equation}}
\newcommand{\ba}{\begin{aligned}} \newcommand{\ea}{\end{aligned}}

\DeclareMathOperator{\Tr}{Tr}

\DeclareRobustCommand\openone{\leavevmode\hbox{\small1\normalsize\kern-.33em1}}%

\newcommand\bigforall{\mbox{\Large $\mathsurround=0pt\forall$}}

\begin{document}

\title{Generalized measurements on qubits in quantum randomness certification and expansion}

\author{Piotr~Mironowicz} \email{piotr.mironowicz@gmail.com}
\affiliation{Department of Physics, Stockholm University, S-10691 Stockholm, Sweden}
\affiliation{Department of Algorithms and System Modeling, Faculty of Electronics, Telecommunications and Informatics, Gda\'nsk University of Technology, Poland}
%\affiliation{International Centre for Theory of Quantum Technologies, University of Gda\'{n}sk, Wita Stwosza 63, 80-308 Gda\'{n}sk, Poland}

\author{Marcus Gr{\"u}nfeld}
\affiliation{Department of Physics, Stockholm University, S-10691 Stockholm, Sweden}

\author{Mohamed Bourennane}
\affiliation{Department of Physics, Stockholm University, S-10691 Stockholm, Sweden}

\date{\today}

\begin{abstract}
Quantum mechanics has greatly impacted our understanding of the microscopic nature. One of the key concepts of this theory is generalized measurements, which have proven useful in various quantum information processing tasks. However, despite their significance, they have not yet been shown empirically to provide an advantage in quantum randomness certification and expansion protocols. This investigation explores scenarios where generalized measurements can yield more than one bit of certified randomness with a single qubit system measurement on untrusted devices and against a quantum adversary. We compare the robustness of several protocols to exhibit the advantage of exploiting generalized measurements. In our analysis of experimental data, we were able to obtain $1.21$ bits of min-entropy from a measurement taken on one qubit of an entangled state. We also obtained $1.07$ bits of min-entropy from an experiment with quantum state preparation and generalized measurement on a single qubit. We also provide finite data analysis for a protocol using generalized measurements and the Entropy Accumulation Theorem. Our exploration demonstrates the potential of generalized measurements to improve the certification of quantum sources of randomness and enhance the security of quantum cryptographic protocols and other areas of quantum information.
\end{abstract}

%\keywords{quantum}

\maketitle

\section{Introduction}

Randomness is a fundamental concept in modern cryptography and information theory~\cite{rfc4086}. Certifying that a source of randomness is unpredictable and unbiased is crucial in many applications. With the rise of quantum technologies, it has become increasingly important to certify quantum sources of randomness, which offer unique advantages in terms of security~\cite{herrero2017quantum,bera2017randomness}.

One approach to quantum randomness certification is based on Bell inequalities, which provide a test for the existence of non-local correlations in a physical system~\cite{bell1964einstein,clauser1969proposed,aspect1982experimental,aspect1982experimental,brunner2014bell}. This provides a self-testing framework, allowing one to check that the device that is being used works properly even if delivered from an untrusted vendor~\cite{mayers1998quantum,pironio2010random,vallone2014quantum,acin2016certified}. Recent work in this field includes employing the Navascués-Pironio-Acin (NPA) hierarchy~\cite{NPA07,NPA08,brown2021device}, and the search for Bell inequalities to efficiently certify quantum randomness~\cite{mironowicz2013robustness,brown2019framework}. Generalized measurements, or Positive Operator-Valued Measures (POVMs), in quantum mechanics, are a powerful tool for information processing~\cite{brandt1999positive,paris2004quantum,hayashi2005asymptotic,lvovsky2009continuous,giovannetti2011advances,zhu2018universally,pezze2018quantum,czerwinski2022selected}. Contrary to projective measurement (PM), they can result in more outcomes than the dimension of the underlying Hilbert space, giving hope to be able to generate more randomness, as shown in~\cite{APVW16,woodhead2020maximal}.

In this work, we present an efficient and practical method for certifying quantum randomness using generalized measurements. Indeed, we have derived a method for self-testing the presence of POVMs~\cite{mironowicz2019experimentally}. We present the certification of more than 1 bit of min-entropy from a single POVM on one of the qubits of an entangled state using a variant of the elegant Bell inequality~\cite{Gisin09}. We also obtain more than 1 bit of randomness in a prepare-and-measure protocol with a similar scenario using a POVM on a single qubit~\cite{mironowicz2019experimentally}. We provide numerical simulations to demonstrate the effectiveness of our randomness certification and expansion.

\section{Methods}

Here, we discuss the POVMs and related concepts in sec.~\ref{sec:povms}. Then in secs~\ref{sec:Bell} and~\ref{sec:pnm} we introduce the basic concepts of Bell non-locality and prepare-and-measure games, which are important as a certificate for the security of protocols of randomness generation. The generated randomness is quantified in terms of so-called min-entropy, and its amount is estimated using the Entropy Accumulation Theorem, as sketched in sec.~\ref{sec:minEAT}. The numerical methods we use in the calculations are described in sec.~\ref{sec:sdp_npa}. In sec.~\ref{sec:CQadversaries} we discuss the difference between classical and quantum adversaries, as both types of eavesdroppers are analyzed throughout this work.

\subsection{Generalized Measurements}
\label{sec:povms}

A crucial tool of this work is properly designed measurement of quantum systems. Generalized measurements allow for a more complete description of quantum systems than the standard PMs by using a broader range of measurement outcomes, thus allowing for more efficient use of the available resources, for instance for the so-called quantum state discrimination~\cite{massar1995optimal,bae2015quantum}. Generalized measurements are essential for achieving the ultimate limits of measurement accuracy in quantum metrology. The use of POVMs has led to significant advancements in quantum metrology, such as the development of optimal quantum sensors and the demonstration of quantum-enhanced precision measurements. These advancements have improved the state of the art in measurement and metrology fundamentals, measurement science, sensors, and measurement instruments~\cite{giovannetti2011advances,toth2014quantum}. Information-complete POVMs, in particular, are sets of quantum measurements that can completely determine the state of a quantum system~\cite{renes2004symmetric,weigert2006simple}. By using information-complete POVMs, one can extract the maximum amount of information about the quantum state being measured, thereby achieving the ultimate limits of measurement accuracy.

A POVM consists of a set of positive semi-definite operators $\{E_i\}_i$ on a Hilbert space $\mathcal{H}$, where the index $i$ labels the possible outcomes of the measurement. The operators satisfy the completeness relation $\sum_i E_i = \openone^{(\mathcal{H})}$, where $\openone^{(\mathcal{H})}$ is the identity operator on $\mathcal{H}$, and the semi-positivity condition $E_i \succeq 0$ for all $i$. The probability of obtaining outcome $j$ in the measurement is given by the Born rule as $p_j = \Tr(\rho E_j)$, where $\rho$ is the density matrix describing the quantum state before the measurement. POVMs can describe a wider range of measurements than PMs, which are a special case of POVMs where each $E_j$ is a projector onto a subspace of $\mathcal{H}$, and for $j \neq j'$ the projectors $E_j$ and $E_{j'}$ are orthogonal.

Any POVM can be realized by initially subjecting a composite system consisting of the original system and a suitably large ancillary system to a unitary transformation, followed by a PM on this combined system. Nonetheless, the unitary transformation needed for this process is typically intricate, and the dimension of the subspace where it has substantial effects grows proportionally with the number of outcomes of the POVM elements. Thus, while PMs are relatively straightforward and commonly used in many experimental setups, POVMs require additional considerations and techniques usually with ancillas~\cite{chen2007ancilla,brida2012ancilla,garcia2021learning,fischer2022ancilla} or quantum random walks~\cite{kurzynski2013quantum,li2019implementation}.

\subsection{Bell Inequalities, Non-Local Correlations and Entanglement}
\label{sec:Bell}

To ensure the security and reliability of quantum information processing, it is essential to have a means of generating and certifying truly random numbers from quantum systems. Bell operators are mathematical expressions that involve the correlations between the outcomes of measurements made on spatially separated quantum systems. The consideration of non-local theories originated with the early works of Einstein, Podolsky, and Rosen~\cite{einstein1935can}. The original Bell operator was proposed by John Bell in 1964~\cite{bell1964einstein} as a means of testing the validity of local hidden variable theories, which posit that the behavior of quantum systems is determined by hidden variables that exist independently of the measurements made on the system.

Experimental tests of the Bell operator have confirmed the existence of non-local correlations in quantum systems, suggesting that quantum systems cannot be described by local hidden variables~\cite{aspect1982experimental,aspect1982experimental2}. Here we consider a bipartite system consisting of two qubits, $A$ and $B$, which are shared between two parties, Alice and Bob, respectively. The parties perform a joint measurement with settings $x$ and $y$ on their qubits and obtain two classical outcomes, $a$ and $b$, respectively. We consider also a third subsystem, called Eve, describing an eavesdropper, and denoted by $E$. Let $\rho^{(ABE)}$ be a joint tripartite quantum system of Alice, Bob, and Eve. The local measurements of Alice and Bob are described by POVMs $\{ \{ M_{a|x} \}_a \}_x$ and $\{ \{ N_{b|y} \}_b \}_y$, respectively. The observed statistics of the measurement outcomes are given by $P(a,b|x,y) = \Tr[\left( M_{a|x} \otimes N_{b|y} \right) \rho^{(ABE)}]$. The correlation functions of Alice and Bob are
\begin{equation}
	\begin{aligned}
		\bkc{A_x}{B_y} \equiv \quad &P(0,0|x,y) + P(1,1|x,y) \\
		&- P(0,1|x,y) - P(1,0|x,y).
	\end{aligned}
\end{equation}
If the observed statistics attain a certain expected value of the Bell operator, which is called a certificate, then we can conclude that the observed outcomes are genuinely random, and not based on any classical correlations between the qubits.

Previous approaches to quantum randomness certification using Bell inequalities have focused on certifying the randomness of binary outcomes from measurements made on separated entangled systems~\cite{pironio2010random,mironowicz2013robustness}. However, these approaches have limitations, since, from outcomes with two possible values, one can obtain at most $1$ bit on min-entropy. A different approach, where entangled qutrits were used to generate $1.106$~bits per round, was taken in~\cite{guo2019experimental}. Unfortunately, the creation of entangled qutrits is more demanding than qubits, and thus attain smaller event rates~\cite{mahmudlu2023fully}. To overcome these limitations, we use the fact that POVMs offer more than two outcomes on a qubit. Let us start by introducing the elegant Bell operator defined by~\cite{Gisin09,APVW16,OBDC18}
\begin{equation}
	\label{eq:el}
	\begin{aligned}
		\beta_{el} = &\bkc{A_1}{B_1} + \bkc{A_1}{B_2} - \bkc{A_1}{B_3} - \bkc{A_1}{B_4} \\
		+ &\bkc{A_2}{B_1} - \bkc{A_2}{B_2}  + \bkc{A_2}{B_3} - \bkc{A_2}{B_4} \\
		+ &\bkc{A_3}{B_1} - \bkc{A_3}{B_2} - \bkc{A_3}{B_3} + \bkc{A_3}{B_4}.
	\end{aligned}
\end{equation}
For local hidden variable theories, $\beta_{el}$ is upper-bounded by $6$, and in quantum theory by $4 \sqrt{3} \approx 6.928$. The latter bound can be achieved only with a unique set of measurements on the maximally entangled state~\cite{smania2020experimental}. The elegant Bell operator does not involve POVMs directly but allows for a construction that can certify their occurrence, as discussed below.

\subsection{Prepare-and-Measure Protocols}
\label{sec:pnm}

Another type of certificate employing measurements on quantum states, but with no need for quantum entanglement, is prepare-and-measure scenarios~\cite{pawlowski2011semi}, where instead of Bell expressions one uses so-called prepare-and-measure games~\cite{brunner2008testing}. In these games, Alice prepares a state $\rho_x$ of a given dimension $d$ depending on her input $x$ and sends it to Bob. Bob performs one of his $m$ measurements, depending on the input $y$, and he obtain $b$ with probability given by $P(b|x,y) \equiv \Tr[\rho_{x} M_{b|y}]$, where $\{M_{b|y}\}_b$ is a POVM in dimension $d$. For given coeffcients $\{\beta_{b,x,y}\}$, a prepare-and-measure game characteristic is given by $\sum_{b,x,y} \beta_{b,x,y} P(b|x,y)$.

An important family of prepare-and-measure games is described by a task called Random Access Codes (RACs). RACs were first introduced in 1983 by S. Wiesner under the name of Conjugate Coding~\cite{wiesner1983conjugate}, and was later reintroduced in~\cite{ambainis1999dense}. An overview of this concept can be found in~\cite{ambainis2008quantum}. In the $m^d \rightarrow 1$ RAC protocol, Alice receives a uniformly distributed string of $m$ dits, $\mathbf{x} = x_1 x_2 \cdots x_{m}$. Her task is to encode this string into a single-dit message $m(\mathbf{x})$ so that Bob can recover any of the $m$ dits with high probability. Bob receives Alice's message $m$, and a referee provides him with a uniformly distributed input $y\in [m]$. Bob then performs classical (possibly probabilistic) computations of some function $b(y, m)$ and outputs $b$. We say that a round of the protocol is successful when it holds that $b(y, m(\mathbf{x})) = x_y$.

The quantum version of RACs is Quantum Random Access Codes (QRACs)~\cite{wiesner1983conjugate,ambainis1999dense}. In the $m^d \rightarrow 1$ QRAC protocol, Alice encodes the $m$-dit input $\mathbf{x}$ into a $d$-dimensional quantum system $\rho_{\mathbf{x}}$, which is afterward transmitted to Bob. Alice's optimal strategy is to send a state $\rho_{\mathbf{x}}$ that maximizes $\Tr \left[ \rho_{\mathbf{x}} \left( \sum_{y \in [m]} M_{x_y|y} \right) \right]$. The average success probability for RACs and QRACs is given by:
\begin{equation}
	\label{eq:psucc}
	\frac{1}{nd^m} \sum_{\mathbf{x} \in [d]^m} \sum_{y \in [m]} P(x_y|\mathbf{x}, y).
\end{equation}
%In recent years, there has been significant progress in the certification of POVMs using dimension witnesses in the prepare-and-measure scenario. Theoretical~\cite{mironowicz2019experimentally} and experimental work~\cite{tavakoli2020self} have been carried out to show that it is possible to certify the dimension of a quantum system by performing measurements on it.

\subsection{Min-entropy and Entropy Accumulation}
\label{sec:minEAT}

Let us introduce now the concept of min-entropy, which is a measure of the amount of uncertainty associated with a probability distribution~\cite{chor1988unbiased,impagliazzo1989pseudo,konig2009operational,issa2017measuring}. For a classical probability distribution $p(x)$, the min-entropy $H_{\infty}(p)$ is defined as the negative logarithm of the largest probability~\cite{chor1988unbiased,impagliazzo1989pseudo,konig2009operational,issa2017measuring} \textit{viz.} $H_{\infty}(p) \equiv -\log\max_x p(x)$. Min-entropy can be interpreted as a measure of the unpredictability of a system and is used in the context of randomness extraction and privacy amplification~\cite{ma2013postprocessing}. A system with high min-entropy is highly unpredictable and thus has strong randomness or privacy guarantees.

The Entropy Accumulation Theorem~\cite{arnon2018practical,dupuis2019entropy,dupuis2020entropy} (EAT) provides a tool for estimating the amount of certified randomness for finite-size data statistics. The min-tradeoff functions to quantify the trade-off between the level of min-entropy and the probability of passing some Bell test, and can be calculated numerically using tools such as the NPA hierarchy discussed in sec.~\ref{sec:sdp_npa}. The resulting certified randomness can improved by passing it through a randomness extractor~\cite{ma2013postprocessing}.

A crucial notion of the EAT is the so-called EAT channels, which are collections of completely positive and trace-preserving maps satisfying certain constraints regarding their input and output systems. In particular, the EAT channels require that in each of the rounds, the inputs presented to the quantum devices used in the protocol are conditionally independent of the previously generated outputs. This constraint is satisfied in all the implementations presented in this paper, as the protocol is assumed to provide the measurement settings from another source of randomness independent of the quantum devices used. This assumption is typical of randomness expansion protocols, where the efficiency of a protocol is defined as the net gain of randomness, so the amount of entropy generated above the amount consumed when running the generator.

Another element needed for the EAT is the min-tradeoff function $f$, which provides a lower bound on the entropy generated in a single round of a generator governed by the laws of quantum mechanics and compatible with a given observed statistics $q$. Let $n$ denote the total number of rounds of the devices in a considered randomness generation process. Let $\vec{A}$ and $\vec{X}$ be strings of length $n$ containing the measurement results provided by Alice's device and the device's inputs, respectively. Let $\vec{B}$ and $\vec{Y}$ be defined analogously for Bob. From these strings, a new string $\vec{Q}$ is calculated element-wise with a function defined by the considered protocol. Let $E$ denote any knowledge possibly possessed by the eavesdropper in a considered scenario. This knowledge can be either classical or quantum, depending on the type of eavesdropper, as elucidated further in sec.~\ref{sec:CQadversaries}. Finally, let $\Omega$ be a set of events $\vec{Q}$ meaning a successful run of the protocol, and the statistics $q$ are calculated from $\vec{Q}$. The EAT provides a bound of the form~\cite{dupuis2019entropy}:
\begin{equation}
	\begin{aligned}
		H^{\epsilon_S}_{\infty} & (\vec{A}, \vec{B} | \vec{X}, \vec{Y}, E, \vec{Q} \in \Omega) \\
		 & \geq n \times \left( \inf_{q \in \Omega} f(q) \right) - \sqrt{n} \times c[\epsilon_S, \Omega, f]
	\end{aligned}
\end{equation}
for a certain functional $c[\cdot,\cdot,\cdot]$. Here $\epsilon_S$ is called the smoothing parameter.

In the so-called spot-checking protocols, the majority of the rounds are generation rounds, where fixed settings of the measurements are used, and their results are used as the outputs containing the obtained random numbers~\cite{coudron2013robust,miller2017universal,brown2019framework}. The other part of the runs are the test rounds used to validate the behavior of the device and its capability of certifying randomness. We denote by $\gamma$ the probability of a test round.

Let $n_X$ and $n_Y$ be the number of possible settings of Alice and Bob, respectively. If Eve has access to the value of Alice's setting $x$ after the randomness is generated, then an upper bound on the guessing probability of the local outcome of Alice is obtained from the following quantity linear in the joint distribution of Alice and Eve, where Eve's outcome is her guess for the outcome of Alice:
\begin{equation}
	\label{eq:guess_prob_local}
	P_{guess}[x] \equiv \sum_{a} P_{AE|X}(a,a|x),
\end{equation}
and $P_{AE|X}(a,e|x)$ is the probability that Alice gets the outcome $a$, Eve the outcome $e$ when Alice uses setting $x$; with Bob's part traced out. The formula~\eqref{eq:guess_prob_local} is relevant to cases where the randomness generation is based on a quantum non-locality in Bell scenarios, with at least a second trusted party, Bob, gathering statistics for certification.

Let us now contrast it with the situation, in which the preparation settings are kept secret. The assumption that Eve does not have access to Alice's setting $x$ is commonly used in the literature~\cite{bowles2014certifying,brask2017megahertz,tebyanian2021semi}. It is relevant to the cases, where the seed used in the certification procedure is not revealed until the generated randomness no longer needs to be concealed, as discussed in Remark~II.5 of~\cite{brown2019framework}. The guessing probability of the local outcome of Bob in this situation is given by the formula
\begin{equation}
	\label{eq:guess_prob_pnm_Avg}
	P_{guess}[y] \equiv \frac{1}{n_X} \sum_{b,x} P_{BE|XY}(b,b|x,y),
\end{equation}
where $P_{BE|XY}$ is the joint conditional probability distribution of the outcome of Bob and Eve, when the state preparation setting of Alice is $x$, and the measurement setting of Bob is $y$. The formula~\eqref{eq:guess_prob_pnm_Avg} is pertinent to prepare-and-measure scenarios, assuming there are multiple different states possible to be prepared by Alice and transmitted to Bob.

The structural similarity of both formulae for guessing probability can be observed. The no-signaling principle states that the marginals of one party do not depend on the setting of the other party, e.g. $P_{A|X}(a|x) = \sum_b P_{AB|XY}(a,b|x,y)$ does not depend on $y$. Because of no-signaling, the expression~\eqref{eq:guess_prob_local} can be rewritten as $\sum_{a,b} P_{ABE|XY}(a,b,a|x,y)$ for arbitrary $y$, or $\sum_{a,b,y} \frac{1}{n_Y} P_{ABE|XY}(a,b,a|x,y)$, where $P_{ABE|XY}$ is the joint conditional probability distribution of Alice, Bob, and Eve. Thus one can consider the summation over unrevealed setting $y$ implicit in~\eqref{eq:guess_prob_local}.
We kept the convention of~\cite{mironowicz2019experimentally,smania2020experimental,tavakoli2020self} for Alice possessing the POVM measurement in the setting with entanglement, and Bob in the prepare-and-measure scenario.

The certified min-entropy under the assumption that Alice and Bob do not use shared variables~\cite{ambainis2008quantum} for a certificate $\mathcal{C}$ attaining the value $q$ is given by
\begin{equation}
	\label{eq:min-entropy}
	\begin{split}
		\text{minimize } &\null \left( -\log_2{P_{guess}} \right) \\
		\text{subject to } &\null \mathcal{C} = q.
	\end{split}
\end{equation}
%To evaluate the certified min-entropy without the assumption of no shared variables, a convex hull of~\eqref{eq:min-entropy} should be taken.

\subsection{Semi-definite Programming and Navascués-Pironio-Acin Technique}
\label{sec:sdp_npa}

The primal optimization task of semi-definite program~(SDP) \cite{vandenberghe1996semidefinite,10.1088/978-0-7503-3343-6,mironowicz2023semi,tavakoli2023semidefinite} is
\begin{align}
	\label{SDP-primal-Boyd}
	\begin{split}
		\text{minimize } &\null c^{T} \cdot x \\
		\text{subject to } &\null F(x) \succeq 0,
	\end{split}
\end{align}
where $F(x)$ is an expression of the form
\begin{equation}
	F(x) \equiv F_0 + \sum_{i \in [m]} x_i F_i,
\end{equation}
$x \in \mathbb{R}^m$ is a variable, $F_i$ (for $i = 0, \cdots, m$) are symmetric constant matrices in $\mathbb{R}^{n \times n}$, and $\succeq 0$ means that a matrix is positive semi-definite. The dual of~\eqref{SDP-primal-Boyd} is
\begin{align}
	\label{SDP-dual-Boyd}
	\begin{split}
		\text{maximize } &\null - \Tr \left[ F_0 Z \right] \\
		\text{subject to } &\null \Tr \left[ F_i Z \right] = c_i, i \in [m],\\
		&\null Z \succeq 0,
	\end{split}
\end{align}
where the variable is a symmetric matrix $Z \in \mathbb{R}^{n \times n}$.

The NPA hierarchy is a sequence of SDPs that approximate the quantum set from the exterior, improving the bounds on quantum correlations with each level of the hierarchy, but at the cost of increased computational resources, as higher levels involve larger variables and additional constraints that tighten the bound~\cite{NPA07,NPA08}. We conducted the numerical calculation using NCPOL2SDPA~\cite{wittek2015algorithm} for modeling NPA, and MOSEK Solver~\cite{mosek21} for solving SDPs. In the NPA context, the SDP matrix, often called the NPA matrix, usually denoted as $\Gamma$, is used to represent and analyze the algebra of observables in quantum systems. The NPA matrix on a bipartite scenario involving Alice and Bob is defined as follows. The entries $\Gamma_{\alpha,\beta}$ correspond to the expectation values of the product of a sequence of operators $\alpha$ and $\beta$ of Alice and Bob, respectively. These operators are typically represented by Hermitian matrices. The entries of the matrix satisfy relations corresponding to certain algebraic properties, including commutation relations, of the relevant sequences of operators. If needed, operators of other parties, including Eve, can also be incorporated into $\Gamma$.

NPA can provide an upper bound on the guessing probability, and thus a lower bound on certified min-entropy against a quantum adversary. To this end, one imposes constraints reflecting the observed statistical description of an experiment $\mathcal{C} =q$ in~\eqref{eq:min-entropy}. As mentioned, one usually considers a violation of a Bell inequality or dimension witness, as a certificate of quantumness. In such a case, the experimental realization consists  of estimation of one parameter, which is later passed, for example to the NPA (see the description below) to determine how much randomness is being certified. In contrast, in some calculations, we utilized a variant of NPA called Nieto-Silleras hierarchy or \textit{more randomness from the same data} method to certify the generated randomness~\cite{bancal2014more,nieto2014using}. This approach takes into account multiple parameters from the experiment.

\subsection{Classical and Quantum Adversaries}
\label{sec:CQadversaries}

In this work, we distinguish between classical and quantum adversaries. The definitions and implications for each type of adversary are detailed below.

A classical adversary can control the process of preparing the quantum devices, overseeing the measurements of Alice and Bob, and creating the entangled state. The correlations between the devices and the classical adversary are assumed to be classical. Specifically, the preparation of the state used by Alice and Bob is given by
\begin{equation}
	\rho^{(AB)} = \sum_e p_e \rho_e^{(AB)},
\end{equation}
where $e$ is a classical variable possessed by the classical adversary. Therefore, the joint state of Alice, Bob, and Eve, is
\begin{equation}
	\rho^{(ABE)} = \sum_e p_e \rho_e^{(AB)} \otimes \proj{e}^{(E)}.
\end{equation}
The eavesdropper's knowledge is represented by the variable $e$. When a particular $\rho_e^{(AB)}$ is fixed, it yields a specific value of the Bell parameter, denoted $q_e$, which may differ from the experimentally observed certificate $\mathcal{C}$ value $q_{\text{exp}}$. The guessing probability conditioned on $e $ is denoted $P_{\text{guess}|e}$. The effective guessing probability is then computed as the convex hull of~\eqref{eq:min-entropy} with
\begin{equation}
	P_{\text{guess}} = \sum_e p_e P_{\text{guess}|e}
\end{equation}
under the constraint that $\mathcal{C} = q_{\text{exp}} = \sum_e p_e q_e$.

A quantum adversary is a more potent entity compared to a classical adversary. The correlations with the eavesdropper are quantum in nature, meaning that the state of the eavesdropper may be entangled arbitrarily with Alice and Bob. Because of quantum steering, the correlations, and consequently the mutual information and guessing probability, between the devices and the eavesdropper can be stronger than in the classical case~\cite{Schroedinger1936,ramanathan2018steering,uola2020quantum}.

When modeling a classical eavesdropper using NPA numerical calculations, the SDP moment matrix involves the operators of Alice and Bob. The SDP optimization is computed separately for various values of the certificate, or in other words, the Bell values. The convex hull of the results is then calculated for the experimentally obtained Bell value. To this end, many optimizations are performed according to~\eqref{eq:min-entropy}. In contrast, to model a quantum eavesdropper in NPA, the SDP moment matrix involves operators of Alice, Bob, and Eve. Consequently, the optimization problem is larger, and thus the SDP method requires more computational power. Fortunately, the calculation is performed only once for the experimentally observed Bell value. Nonetheless, since SDP solving does not scale linearly with the size of the problem, modeling the quantum eavesdropper is computationally more demanding.

\section{Results}

In sec.~\ref{sec:setup} we describe the certificates we use. Then, in sec.~\ref{sec:more1bit_from_qubit} we show how in the asymptotic case we can certify more than $1$ bit from a single (local) measurement on a qubit in a scenario with Bell non-locality as a certificate. The global randomness obtained by taking into account the results of both Alice and Bob is evaluated in sec.~\ref{sec:global_randomness_entangled}. The prepare-and-measure results are in sec.~\ref{sec:povmsPnM}. We compare the efficiency of the proposed protocols with other protocols not involving POVMs in sec.~\ref{sec:compare_with_PMs}. Finally, in sec.~\ref{sec:finite_data}, we use the EAT to calculate the amount of randomness that can be obtained in a randomness expansion if the proposed Bell certificate was used.

\subsection{Experiments Involving Certification of Generalized Measurements}
\label{sec:setup}

We consider two protocols for randomness generation, \textit{viz.} a protocol working in a scenario with Bell non-locality and guessing probability defined by~\eqref{eq:guess_prob_local}, and a protocol in prepare-and-measure setup with guessing probability given by~\eqref{eq:guess_prob_pnm_Avg}. For the former, we have chosen the scheme with the elegant Bell operator~\eqref{eq:el}, and for the latter, we consider the reduced $3$ to $1$ QRAC, as discussed below; see~\eqref{eq:reduced}. The reduced $3$ to $1$ QRAC has four preparation states, so the same number as the full $2$ to $1$ QRAC. Next, we conduct a numerical assessment of the randomness produced by previous experiments conducted in 2019~\cite{smania2020experimental,tavakoli2020self}. We present the experimental data yielding more than $1$ bit of randomness from a single-qubit measurement using a couple of statistical parameters as a certificate and then analyze the robustness of randomness generation techniques using single parameter certificates.

The first of the analyzed experiments focused on the conditional probabilities in the Bell expression given by the following equation:
\begin{equation}
	\label{eq:certificate_Entangled}
	\beta^k_{el}=\beta_{el} - k \sum^4_{i=1} P(a=i,b=+1|x=4,y=i),
\end{equation}
with $k > 0$ and $\beta_{el}$ given by~\eqref{eq:el}. The value of the Bell operator $\beta^k_{el}$ allowed by quantum mechanics is upper-bounded by $4\sqrt{3}$, the same as the original expression in~\eqref{eq:el}. This maximum can be reached only with the maximally entangled two-qubit state, and the observables $\bar{A}_1 = \sigma_x$, $\bar{A}_2 = \sigma_y$ and $\bar{A}_3 = \sigma_z$ for Alice, and $\bar{B}_1 = \frac{1}{\sqrt{3}}(\sigma_x-\sigma_y+\sigma_z)$, $\bar{B}_2 = \frac{1}{\sqrt{3}}(\sigma_x+\sigma_y-\sigma_z)$, $\bar{B}_3 = \frac{1}{\sqrt{3}}(-\sigma_x-\sigma_y-\sigma_z)$ and $\bar{B}_4 = \frac{1}{\sqrt{3}}(-\sigma_x+\sigma_y+\sigma_z)$ for Bob.

The second term in~\eqref{eq:certificate_Entangled} is equal to zero only if $\bar{A}_4$ is an information-complete POVM on a qubit, whose components are anti-aligned with Bob's measurements $\bar{B}_y$. In other words, $\bar{A}_4 = \{ M_{a|4} \}_a$ is the four-outcome POVM, consisting of the following elements:
\begin{equation}
	\label{eq:povm}
	\begin{aligned}
		M_{1|4}&=\frac{1}{2}\begin{bmatrix} \alpha&-\beta(1+i)\\ \beta(-1+i)&1-\alpha \end{bmatrix},\\
		M_{2|4}&=\frac{1}{2}\begin{bmatrix} 1-\alpha&\beta(-1+i)\\ -\beta(1+i)&\alpha \end{bmatrix},\\
		M_{3|4}&=\frac{1}{2}\begin{bmatrix} 1-\alpha&\beta(1-i)\\ \beta(1+i)&\alpha \end{bmatrix},\\
		M_{4|4}&= \frac{1}{2}\begin{bmatrix} \alpha&\beta(1+i)\\ \beta(1-i)&1-\alpha \end{bmatrix},
	\end{aligned}
\end{equation}
where $\alpha= \frac{3-\sqrt{3}}{6}$ and $\beta= \frac{\sqrt{3}}{6}$. In this case, the four unit vectors associated with the components of $\{ M_{a|4} \}_a$ define a regular tetrahedron within the Bloch sphere.  We note that the POVM~\eqref{eq:povm} is information-complete~\cite{smania2020experimental}.

In the prepare-and-measure case, we adapt the results of the experiment~\cite{tavakoli2020self}. The certificate is derived from~\eqref{eq:certificate_Entangled} in~\cite{mironowicz2019experimentally} using the method of Mironowicz-Li-Paw{\l}owski (MLP)~\cite{mironowicz2014properties} and is given by the following expression of a prepare-and-measure task, which is not a standard QRAC, but a reduced $3$ to $1$ QRAC:
\begin{equation}
	\label{eq:reduced}
	R_{3 \to 1} \equiv \frac{1}{12} \sum_{\mathbf{x} \in \{\tilde{1},\tilde{2},\tilde{3},\tilde{4}\}} \sum_{y \in [3]} P(x_y|\mathbf{x}, y),
\end{equation}
where we identify $\tilde{1} \equiv 000$, $\tilde{2} \equiv 011$, $\tilde{3} \equiv 101$, and $\tilde{4} \equiv 110$. In the perfect case the success probability of~\eqref{eq:reduced} is
\begin{equation}
	\label{eq:succ3to1Value}
	S_3 = \frac{1}{2} \left(1 + \frac{\sqrt{3}}{3}\right) \approx 0.78868.
\end{equation}
Similar to the case with the elegant Bell expression with additional terms~\eqref{eq:certificate_Entangled}, we introduce an additional fourth input of Bob, related to a four-outcome measurement (with outcomes labeled $1,2,3,4$):
\begin{equation}
	\label{eq:certificate_PnM}
	R_{3 \to 1} - k \times [P(1|\tilde{1},4) + P(2|\tilde{2},4) + P(3|\tilde{3},4) + P(4|\tilde{4},4)]
\end{equation}
for some $k > 0$. The expression~\eqref{eq:certificate_PnM} cannot have a value greater than $S_3$, and that value would be obtained only when the states and measurements optimal for~\eqref{eq:succ3to1Value} are used, and the second part of~\eqref{eq:certificate_PnM} is equal to zero; these imply that the fourth measurement is a POVM given by~\eqref{eq:povm}, as shown in~\cite{mironowicz2019experimentally}. In the noiseless case, each of the three results of the POVM measurement non-orthogonal to the prepared state occurs with the same probability. The orthogonal result has the probability of occurrence equal to zero. Thus, the maximal randomness possible to be certified with this setup using the POVM is $-\log_2{\frac{1}{3}} \approx 1.585$.

\subsection{Certification of more than 1 Bit of Randomness from a Local Measurement on a Single Qubit with Elegant Bell Operator secure against a Quantum Adversary}
\label{sec:more1bit_from_qubit}

\begin{table}[htbp]
	\begin{center}
		\small
		\begin{tabular}{|l|l|l|}
			\hline
			Setting & Theory & Experiment \\ \hline
			$\bkc{A_1}{B_1}$ & $1/\sqrt{3}\approx0.577$ & $0.553 \pm 0.002$ \\
			$\bkc{A_1}{B_2}$ & $0.577$ & $0.573 \pm 0.002$ \\
			$\bkc{A_1}{B_3}$ & $-0.577$ & $-0.581 \pm 0.002$ \\
			$\bkc{A_1}{B_4}$ & $-0.577$ & $-0.543 \pm 0.002$ \\
			$\bkc{A_2}{B_1}$ & $0.577$ & $0.589 \pm 0.002$ \\
			$\bkc{A_2}{B_2}$ & $-0.577$ & $-0.599 \pm 0.002$ \\
			$\bkc{A_2}{B_3}$ & $0.577$ & $0.529 \pm 0.002$ \\
			$\bkc{A_2}{B_4}$ & $-0.577$ & $-0.579 \pm 0.002$ \\
			$\bkc{A_3}{B_1}$ & $0.577$ & $0.584 \pm 0.002$ \\
			$\bkc{A_3}{B_2}$ & $-0.577$ & $-0.557 \pm 0.002$ \\
			$\bkc{A_3}{B_3}$ & $-0.577$ & $-0.621 \pm 0.002$ \\
			$\bkc{A_3}{B_4}$ & $0.577$ & $0.601 \pm 0.002$ \\
			$\beta_\text{el}$ & $4 \sqrt{3}\approx6.928$ & $6.909 \pm 0.007$ \\
			$P(1,+|4,1)$ & $0$ & $0.0021 \pm 0.0001$ \\
			$P(2,+|4,2)$ & $0$ & $0.0020 \pm 0.0001$ \\
			$P(3,+|4,3)$ & $0$ & $0.0025 \pm 0.0001$ \\
			$P(4,+|4,4)$ & $0$ & $0.0025 \pm 0.0001$ \\ \hline
		\end{tabular}
	\end{center}
	\caption{\label{tab:proj} Experimental values for the combinations of settings needed to test the elegant Bell expression. The experimental data are taken from~\cite{smania2020experimental}.}
\end{table}

\begin{figure}[htbp]
	\begin{tikzpicture}
		% Outer ellipse
		\draw (0,0) ellipse (4cm and 2cm);
		\node at (0,-0.6) {$\mathcal{Q}_{=}$};
		
		% Middle ellipse
		\draw (0,0) ellipse (3cm and 1.5cm);
		\node at (0,-1.2) {$\mathcal{Q}_{2+ABE, =}$};
		
		% Inner ellipse
		\draw (0,0) ellipse (2cm and 1cm);
		\node at (0,-1.8) {$\mathcal{Q}_{2+ABE, \geq}$};
	\end{tikzpicture}
	\caption{Illustration of the relaxations of the quantum set of probability distributions under the constraint that the probability distributions have correlators exactly equal to those obtained in the experiment, referred to as \textit{equality constraining}. This set is denoted as $\mathcal{Q}_{=}$. This set is contained in a relaxed set $\mathcal{Q}_{2+ABE, =}$ obtained at the level $2+ABE$ of NPA with the same equality constraining. Further relaxation is obtained if the equalities are replaced with inequalities, denoted as $\mathcal{Q}_{2+ABE, \geq}$. Any set satisfying the equality constraining satisfies the inequalities.}
	\label{fig:SDPrelax}
\end{figure}
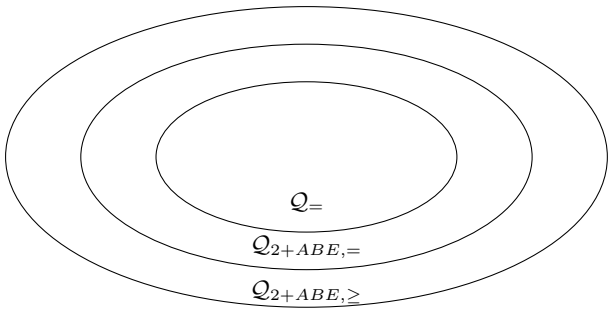

For the experiment involving quantum entanglement, we focus on calculating the randomness of the single-party measurement with the POVM of Alice. We used the Nieto-Silleras technique at the NPA level $2+ABE$ with the following constraints. We took the values of correlators from~\cite{smania2020experimental}, as given in Tab.~\ref{tab:proj}. It can be seen that the POVM term from~\eqref{eq:certificate_Entangled} is $-k \cdot 0.0091$. The most natural assumption regarding correlators would be to optimize with the constraint that the probability distributions have correlators exactly equal to those obtained in the experiment. Instead, we relaxed this constraint and enforced only that positive correlators are at least as large and negative correlators are at most as large as those obtained from the experiment, as illustrated in Fig.~\ref{fig:SDPrelax}. The use of relaxations of the set of quantum probabilities is a standard practice in cryptography, providing security at least as strong as if optimization was performed on a non-relaxed exact quantum set, since it not only allows an eavesdropper to use any attacks allowed by quantum mechanics but also includes hypothetical additional attacks possible in the relaxed set~\cite{pironio2010random,acin2016certified,vazirani2019fully,pirandola2020advances,liu2022toward}.

In these calculations of the certified randomness, we assume that Eve has access to the value of $x$ after the randomness is generated. The guessing probability~\eqref{eq:guess_prob_local} obtained for the local POVM for the setting $x = 4$ of Alice is equal to $0.4325$, which refers to the min-entropy $-\log_2(0.4325) \approx 1.21$~bits secure against a quantum adversary. The dual of SDP provides the following new Bell expression, which serves as a certificate of randomness:
\begin{equation}
	\begin{aligned}
		20.391 + & 2.711 \bkc{A_1}{B_1} + 2.936 \bkc{A_1}{B_2} - 2.845 \bkc{A_1}{B_3} \\
		&- 2.813 \bkc{A_1}{B_4} + 2.925 \bkc{A_2}{B_1} - 3.031 \bkc{A_2}{B_2} \\
		&+ 2.622 \bkc{A_2}{B_3} - 2.954 \bkc{A_2}{B_4} + 2.837 \bkc{A_3}{B_1} \\
		&- 2.866 \bkc{A_3}{B_2} - 3.051 \bkc{A_3}{B_3} + 3.079 \bkc{A_3}{B_4} \\
		&- 4.104 P(1,+1|4,1) - 4.517 P(2,+1|4,2) \\
		& - 3.847 P(3,+1|4,3) - 4.203 P(4,+1|4,4).
	\end{aligned}
\end{equation}
The amount of the certified min-entropy is similar to that obtained in the maximal violation of the Clauser-Horne-Shimony-Holt (CHSH) inequality, but therein two-qubit measurement is required. Here we measure only a single-qubit system. We note that, for the randomness of the expression~\eqref{eq:guess_prob_local} Bob does not need to perform a measurement, and in the noiseless case, each of the outcomes has the same probability giving up to $-\log_2{\frac{1}{4}} = 2$ bits of certified randomness. We exploit that Bob's part is used only for test rounds further in sec.~\ref{sec:finite_data} to propose a measurement-device-independent (MDI) randomness expansion protocol.

\subsection{Two-Party Randomness from Generalized Measurements in Elegant Bell Expression secure against a Classical Adversary}
\label{sec:global_randomness_entangled}

To calculate the global randomness certified in the experiment, we used the ordinary elegant Bell operator~\eqref{eq:certificate_Entangled} and the method of~\cite{Armin16}. This means that the guessing probability is averaged over four settings of Bob, \textit{i.e.} we consider the following formula:
\begin{equation}
	\label{eq:guess_prob_Entangled}
	P_{guess} \equiv \max_{\substack{a \in \{0,1,2,3\}^4 \\ b \in \{0,1\}^4}} \frac{1}{4} \sum_{y=1,2,3,4} P(a_y, b_y|x=4, y).
\end{equation}
In formula~\eqref{eq:guess_prob_Entangled}, we consider all possible deterministic guessing strategies of the pair of outcomes $(a,b)$ by Eve when she knows the measurement setting $y$ of Bob. The probabilistic strategies are obtained by taking the convex hull of the guessing probability plot over possible values of the Bell expression. Each value of $x$ requires $4^4 \times 2^4 = 4096$ semi-definite optimizations to calculate~\eqref{eq:guess_prob_Entangled}.

As the level of the NPA hierarchy increases, the execution time required for the calculations grows significantly. Therefore, it is important to find a way to obtain accurate bounds while minimizing the computational cost. Let $l$ denote the desired level of the NPA hierarchy (for the sake of simplicity we do not consider the intermediate levels). One way to achieve this is to focus on the most probable outcomes of the measurement. Since the value of the guessing probability is determined by the probability of the most probable outcome, it is not necessary to bound exactly the probabilities of the less probable outcomes. Instead, one can iteratively increase the level of the NPA hierarchy but only apply it to the most probable outcomes. Thus, we get the bound on the guessing probability with a given level $l$ of the hierarchy saving calculation time. We applied this approach both in entanglement and, further in this work (see sec.~\ref{sec:povmsPnM}) for the prepare-and-measure certification. More precisely, we use the following method for efficient guessing probability calculation in the Bell expression (or, further in sec.~\ref{sec:povmsPnM}, in prepare-and-measure game). First, we set
\begin{equation}
	\bigforall_{\substack{a \in \{0,1,2,3\}^4 \\ b \in \{0,1\}^4}} \gamma(a,b) = 1, \lambda(a,b) = 0,
\end{equation}
or, for the prepare-and-measure case considered further in sec.~\ref{sec:povmsPnM}
\begin{equation}
	\bigforall_{b \in \{0,1,2,3\}^4} \gamma(b) = 1, \lambda(b) = 0.
\end{equation}
Then we follow these steps:
\begin{enumerate}
	\item Find $(a',b')$ (or $b'$) such that $\gamma(a',b')$ (or $\gamma(b')$) attains the largest value over $\gamma$.
	\item If $l' = \lambda(a'b') < l$ (or $l' = \lambda(b') < l$) then perform optimization at the level $l' + 1$ and set $\gamma(a',b')$ (or $\gamma(b')$) to the resulting value of the optimization, and $\lambda(a'b') = l' + 1$ (or $\lambda(b') = l' + 1$). Go to step 1.
	\item If $\lambda(a'b') = l$ (or $\lambda(b') = l$) then the guessing probability is $\gamma(a',b')$ (or $\gamma(b')$); finish the calculations.
\end{enumerate}

\begin{figure}[htbp]
	\includegraphics[width=\linewidth]{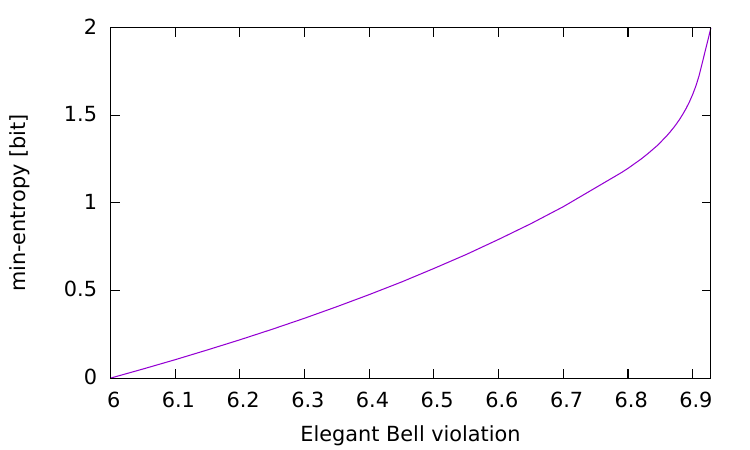}
	\caption{The min-entropy certified by given experimental values of modified elegant Bell operator~\eqref{eq:certificate_Entangled} with $k = 2$. For the Bell value obtained in the experiment~\cite{smania2020experimental}, equal to $6.8907$, the certified min-entropy is $1.55$ bits secure against a classical adversary. In the perfect case, 2 bits of min-entropy are certified.}
	\label{fig:randomness_plot}
\end{figure}

The randomness plot with the figure of merit~\eqref{eq:guess_prob_Entangled} for different values of the certificate~\eqref{eq:certificate_Entangled} is shown in Fig.~\ref{fig:randomness_plot}. The function is convex. In particular, the value of~\eqref{eq:certificate_Entangled} obtained in the experiment is
$6.8907$ % 6.890753
certifying $1.55$ bits of min-entropy from a two-qubit system secure against a classical adversary.

\subsection{Randomness from Generalized Measurements in the Prepare-and-Measure Protocol}
\label{sec:povmsPnM}

Similarly as in sec.~\ref{sec:global_randomness_entangled}, to calculate the guessing probability for the prepare-and-measure protocol, we average over four settings of state preparations of Alice and fix the measurement of Bob to be $y=4$. The guessing probability for this case is
\begin{equation}
	\label{eq:guess_prob_PnM}
	P_{guess} \equiv \max_{b \in \{0,1,2,3\}^4} \frac{1}{4} \sum_{x=1,2,3,4} P(b_x|x, y=4).
\end{equation}
In formula~\eqref{eq:guess_prob_PnM} we consider all possible deterministic guessing strategies of the outcome $b$ by Eve when she knows the preparation setting $x$ of Alice. Similarly, as in the previous situation, the probabilistic strategies are obtained by taking the convex hull of the guessing probability plot over possible values of the prepare-and-measure game, as explained in sec.~\ref{sec:CQadversaries}.

\begin{figure}[htbp]
	\centering
	\includegraphics[width=\linewidth]{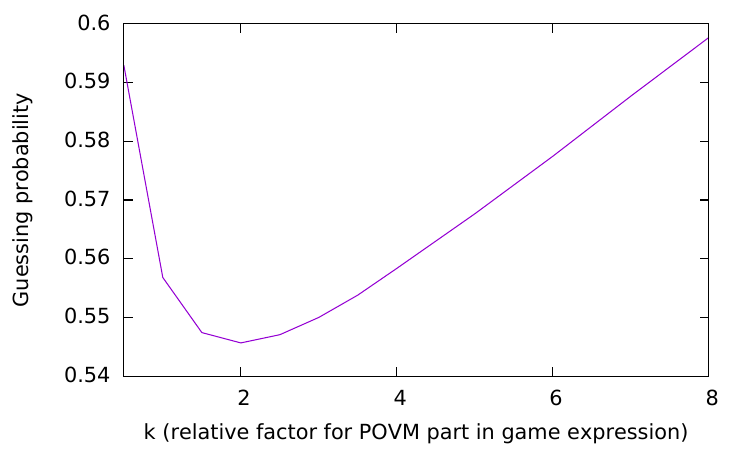}
	\caption{The dependence of the guessing probability on the coefficient $k$ in the additional term of the prepare-and-measure protocol. The optimal value of $k$ is $2$, which gives a guessing probability of $0.54565$. The guessing probability is calculated for the Bell expression~\eqref{eq:certificate_PnM} using the data from the experiment~\cite{smania2020experimental}. The values are obtained using level $3$ of MLP.}
	\label{fig:experimentalrandomnesspnmdifferentk}
\end{figure}

We used experimental data from our work~\cite{tavakoli2020self} and analyzed the data using our method of~\cite{mironowicz2014properties} with a prepare-and-measure game~\eqref{eq:certificate_PnM} containing an additional term with a coefficient $k$. Fig.~\ref{fig:experimentalrandomnesspnmdifferentk} shows the guessing probability~\eqref{eq:guess_prob_PnM} as a function of the coefficient $k$ in the certificate~\eqref{eq:certificate_PnM}. For each value of $k$, we recalculated the value of the certificate attained with the experimental data~\cite{smania2020experimental}. Then, we constrained the optimization~\eqref{eq:min-entropy} with this value of the certificate and optimized for the guessing probability given by~\eqref{eq:guess_prob_PnM}. Fig.~\ref{fig:experimentalrandomnesspnmdifferentk} shows that, for the experimental data, the most randomness is certified with $k = 2$. The results demonstrate the importance of carefully choosing the coefficients in the certificate to maximize the generated randomness.

\begin{figure}[htbp]
	\centering
	\includegraphics[width=\linewidth]{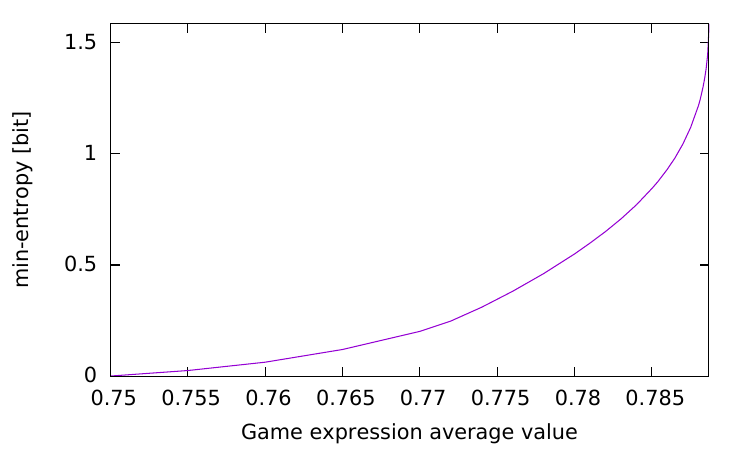}
	\caption{Demonstration of the robustness of min-entropy certification with the prepare-and-measure protocol when an additional term is added with the coefficient $k=2$, which is shown to be optimal. The values are obtained using level $3$ of MLP. The maximal value is $\log_2 (1/3) \approx 1.58$. The experiment~\cite{tavakoli2020self} attained a value of the certificate~\eqref{eq:certificate_PnM} equal to $0.785361$ giving $0.87$ bits of min-entropy. The critical value of the certificate that allows certifying more than 1 bit of min-entropy is $0.787$.}
	\label{fig:randomnessplotpnm}
\end{figure}

Fig.~\ref{fig:randomnessplotpnm} shows the robustness of min-entropy certification using~\eqref{eq:certificate_PnM} with the coefficient $k = 2$. The guessing probability curve is convex, thus the assumption of no shared variables is not needed. The data from the experiment~\cite{tavakoli2020self} lead to a guessing probability of $0.54565$ and certify $-\log_2(0.54565) \approx 0.87$~bits of randomness in terms of min-entropy, and the certificate~\eqref{eq:certificate_PnM} equals $0.785361$ secure against a classical adversary when the measurement settings are revealed.

Next, we used experimental data from our work~\cite{tavakoli2020self}, which employed a prepare-and-measure game containing the additional term, \textit{i.e.}~\eqref{eq:certificate_PnM}, and assumed the asymptotic case with no uncertainty of statistics, and again followed the Nieto-Silleras constraining. We analyze the data using MLP. Whereas, in the former experiment, we considered the case when the setting $x$ is revealed to Eve, in this experiment we assume it to be secret and unknown to Eve, and thus for the guessing probability upper bound we use the formula~\eqref{eq:guess_prob_pnm_Avg}, with $n_X = 4$. We fixed here the setting to be the POVM, \textit{viz.} $y=3$. We have executed the MLP with the level $2+AAE+BBE+ABE$ and dimension two. We imposed the Nieto-Silleras-type constraints on the observed events~\cite{nieto2014using,bancal2014more}. The guessing probability we obtained is equal to $0.4771$, meaning $1.07$ bits of min-entropy secure against a quantum adversary when the measurement settings are kept secret.

\subsection{Comparison of Generalized Measurements and Projective Measurements for Randomness Certification Protocols}
\label{sec:compare_with_PMs}

We now compare the efficiency of randomness generation using POVMs and PMs. To this end, we juxtapose the protocols exploiting POVMs discussed above with the following protocols employing PMs. For the Bell non-locality, we choose as a reference the well-known CHSH~\cite{clauser1969proposed}. For the prepare-and-measure scenario, we consider the protocol based on the modified CHSH introduced in~\cite[eq.(28)]{mironowicz2014properties}. We calculated the robustness of the compared protocols using NPA and MLP, both at level $2+ABE$, for Bell non-locality certificates, and prepare-and-measure certificates, respectively~\cite{NPA07,NPA08,mironowicz2014properties}. For each certificate we denote by $q$ the considered value, \textit{e.g.} the one obtained in an experiment, by $T$ the Tsirelson bound of the certificate, \textit{i.e.} its maximal value possible in quantum mechanics, and by $W$ the value of the certificate attained on the probability distribution with all outcomes distributed uniformly. To express the noise we define the relative certificate value as $\eta \equiv \frac{q-W}{T-W}$.

In the experiment with entangled states~\cite{smania2020experimental}, the elegant Bell operator part obtained the value $6.908753$, as can be seen from Tab.~\ref{tab:proj}, and the POVM term obtained the value $-k \cdot 0.0091$. We determine the relative violation $\eta^{ent}$ for this case as follows. For the uniform distribution of outcomes all correlators vanish, and thus in the Bell operator~\eqref{eq:certificate_Entangled} the white noise value is $W^{ent} = -k \cdot 4 \cdot \frac{1}{8} = - k / 2$, and thus for $k=1$ we get $\eta^{ent} = 0.9961565$. % entangledRandomnessPlot4outcomes.py, line 24

Similarly, for the prepare-and-measure scenario~\cite{tavakoli2020self}, the white noise value is calculated as $W^{pnm} = 0.5 - k / 4$. For the certificate~\eqref{eq:certificate_PnM} we obtained in the experiment: $R_{3 \to 1} = 0.786488$, $P(1|\tilde{1},4) = 0.00190204$, $P(2|\tilde{2},4) = 0.00109037$, $P(3|\tilde{3},4) = 0.00164225$, $P(4|\tilde{4},4) = 0.00212785$. Thus for $k=1$ we get $\eta^{pnm} = 0.9833858$. % pnmRandomnessPlot4outcomes.py line 25

\begin{figure}[htbp]
	\centering
	\includegraphics[width=0.95\linewidth]{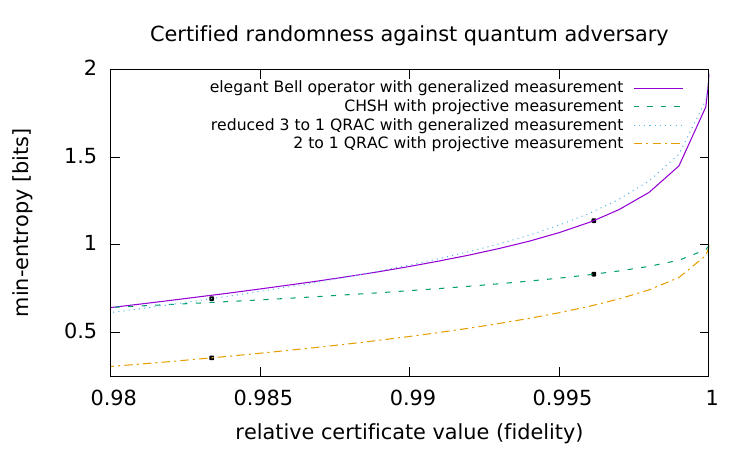}
	\caption{(color online) Robustness of randomness certification protocols using various certificates. The black dots refer to the relative certificate values obtained in experiments from~\cite{smania2020experimental,tavakoli2020self}, \textit{viz.} $\eta^{ent} = 0.9962$ and $\eta^{pnm} = 0.9834$ for Bell non-locality certificate~\eqref{eq:certificate_Entangled} with $k = 1$, and prepare-and-measure certificate~\eqref{eq:certificate_PnM} with $k = 1$, respectively. The other experiment by Liu \textit{et al.}~\cite{liu2019experimental} exploiting elegant Bell operator achieved $\eta^{liu} = {6.8138} / {4 \sqrt{3}} \approx 0.9835$ was not able to exhibit any advantage of POVMs, as for this relative certificate value for Bell non-locality protocols, the efficiency of that operator and CHSH with PMs are similar; it also lacked certification of the POVM form (cf.~\cite{smania2020experimental}) and thus was not device-independent.}
	\label{fig:robustnessquantum}
\end{figure}

The consequences of these calculations for the amount of the certified randomness for each case are shown in Fig.~\ref{fig:robustnessquantum}. The protocols using POVM certify $2$ bits of min-entropy in the perfect case $\eta = 1$, whereas the protocols with PMs certify then only $1$ bit of min-entropy. It can be seen that for the values of $\eta$ acquired in state-of-the-art experiments, the protocols utilizing POVMs are much more robust than those employing PMs. The values in the figures are derived with a single parameter certificate, expressed by~$\eta$.

\subsection{Finite Data Analysis of Randomness Expansion with Generalized Measurements in Entanglement Scenario}
\label{sec:finite_data}

Any implementation of a randomness generator using the outcome averaged over a couple of settings, as is the case of~\eqref{eq:guess_prob_Entangled} and~\eqref{eq:guess_prob_PnM}, itself requires random bits. As in these formulae for the guessing probabilities we cover the case when the setting is known by Eve, these bits can come from a public source. It is only needed that they are not accessible to the other part of the device before the outcome has been generated. In the previous considerations, we considered the asymptotic cases, with no uncertainty of the value of a particular certificate. In real-world implementations of practical randomness expansion, finite data statistics have to be considered instead. Here we provide such an analysis.

To perform the analysis using the EAT we need to evaluate specific quantities describing the protocols. The experiment reported in~\cite{smania2020experimental} achieved the relative certificate value of $\eta^{ent} = 0.9962$ with a standard deviation of about $0.001$. Numerical differentiation with the forward difference method at the point $\eta = 0.993$ gives a tangent line of a form $f(\eta) \equiv 42.07 \eta -40.797$ [bits per round], serving as the lower bound on the randomness, i.e. the min-tradeoff function. The algebraic upper bound on~\eqref{eq:certificate_Entangled} is $12$, and the quantum lower bound is $- 4 \sqrt{3} - 1$, thus $\eta \in [-1 - \sqrt{3}/12, \sqrt{3}]$, and the range of values of $f$ is $D \equiv f(\sqrt{3}) - f(-1 -\sqrt{3}/12) \approx 121.01$.

We follow the formulation of the EAT given in Theorem II.1 of~\cite{brown2019framework}. For $\beta \in (0,1)$ we define
\begin{subequations}
	\label{eqs:EAT_functions}
	\begin{equation}
		\epsilon_V(\beta, \gamma) \equiv \frac{\beta \cdot \ln(2)}{2} \cdot \left[ \log_2(129) + \sqrt{D^2 / \gamma + 2} \right]^2,
	\end{equation}
	\begin{equation}
		\epsilon_K(\beta) \equiv \frac{\beta^2 \cdot \left( 2^{\beta \cdot \left(3 + D\right)} \right)}{6 (1-\beta)^3 \cdot \ln(2)} \cdot [\ln(2) \cdot (2 + D) + 2]^3,
	\end{equation}
	\begin{equation}
		\epsilon_\Omega(\beta, p_\Omega, \epsilon_S) \equiv (1 - \log_2(p_\Omega \cdot \epsilon_S)) / \beta.
	\end{equation}
\end{subequations}
For three standard deviations confidence interval, assuming that the Bell expression value distribution is well approximated with the normal distribution, we have the probability of the successful run of the protocol given by $p_\Omega = 0.997$. We follow the high-quality randomness expansion of~\cite{liu2021device} and choose the value $\epsilon_S = 3.09 \times 10^{-12}$. To analyze the efficiency of a protocol utilizing the presented method of certification using POVMs, we need to find relevant values of the protocol parameters $\beta$ and $\gamma$. The former parameter is arbitrary, and the latter refers to the probability of a test round, as described in sec.~\ref{sec:minEAT}.

\section{Discussion}

The randomness expansion considered in this work belongs to the device-independent (DI) class, see secs~\ref{sec:more1bit_from_qubit} and~\ref{sec:global_randomness_entangled}; or the semi-device-independent (semi-DI) class with dimension constraint~\cite{10.5555/2011827.2011830,pawlowski2011semi} (see sec.~\ref{sec:povmsPnM}).

Generating quantum randomness when the devices are untrusted is much more difficult compared to trustworthy equipment~\cite{mannalath2022comprehensive}.
Trusted quantum devices with $100$~Gbps (gigabits per second) of randomness have been reported~\cite{bruynsteen2023100,luo2023recent}.
The Source-device-independent (source-DI) protocols where the measurements are considered as trusted can reach $17 Gbps$~\cite{avesani2018source}.
%The Measurement-DI protocols have much lower generation rates of about $5$~Kbps. This indicates that the measurement certification in a device-independent way is more demanding than source certification.
The highest rate achieved to date with fully DI-quantum random number generators (QRNGs) is $13$~Kbps~\cite{liu2021device}.

The proposed solution provides a step towards overcoming the limits of fully DI-QRNG, which was possible due to the exploitation of increased generation rate per event obtained with POVMs instead of the PMs utilized in traditional QRNG approaches. A key advantage of using POVMs is the increase in the amount of generated raw bits per event.
%In the past increasing the rate per round was providing a significant boost in generation rates of randomness in various classes of quantum protocols.
%For instance, the most efficient Semi-DI-QRNG with dimension or energy constraints was realized in 2017~\cite{brask2017megahertz}. It achieved the generation rate of $16.5 Mbps$ and $0.33$ bits per round. On the other hand, the recent developments in Semi-DI-QRNG have smaller generation rates but concentrate on developing solutions predicted to scale better with future technologies. The work~\cite{leone2022certified} announced in 2022 achieves only $4.4$~Kbps of randomness with $0.025$ bit per round. We expect that implementing our protocol with the novel technologies is a reasonable direction for advancing DI-QRNGs.
A significant gain in the generation rate of randomness with a higher rate per event is also obtained from saving the seed randomness needed at the randomness extraction stage~\cite{ma2013postprocessing}.
To better exhibit its benefits, below we relate our protocols with other existing solutions. 

\paragraph{Device independent protocols.}
In this paper, we present conceptual progress in quantum randomness certification by demonstrating that it is possible to obtain more than a bit of randomness from a single generalized measurement on a qubit system certified by a violation of a Bell inequality. Recall that we obtained $1.21$~and $1.55$~bits of min-entropy with the use of the generalized measurements on single- or two-qubit systems, respectively, in full DI security.

The closest work is Liu \textit{et al.} reporting the elegant Bell operator experimental value of $6.8138$, which refers to the relative certificate value $\eta^{liu} = 0.9835$, as defined in sec.~\ref{sec:compare_with_PMs}~\cite{liu2019experimental,liu2020experimental}. As can be seen from Fig.~\ref{fig:robustnessquantum}, for this value the gain from using POVM instead of PMs for QRNG with entanglement-based protocols is negligible and does not compensate for the effort of constructing a more complicated setup realizing the more complex measurements. In their paper, the authors gathered $74802774$ raw bits in 4 hours (meaning about $2.6$k of events per second). Nonetheless, the work~\cite{liu2019experimental} is does not perform a DI security analysis and only concludes that the bits are close to fully random from the observation that the observed value of the Bell certificate is relatively high. As can be seen from Fig.~\ref{fig:randomness_plot}, the Bell value announced in that work yields about $1.25$~bits per round, thus one can expect about $3.25$k raw bits per second in the asymptotic case.

%\paragraph{Semi-Device independent.}
%We compared these results with the prepare-and-measure scenario, which is usually considered more efficient at the price of the assumption regarding the dimension of the transferred quantum system. Surprisingly, with this scenario using an experimental setup of similar quality~\cite{tavakoli2020self}, the randomness obtained with generalized measurements was significantly slower, \textit{viz.} about $0.87$ bits of min-entropy when the preparation settings are revealed or $1.07$ bits if they are kept secret, with Semi-DI security.
%This result is particularly valuable since randomness is a critical resource in many applications of quantum information theory, such as quantum cryptography and quantum key distribution.

\paragraph{Measurement-Device independent protocols.}
Let us assume that the test rounds are performed using the certificate~\eqref{eq:certificate_Entangled} with $k = 1$, whereas the generation rounds employ a trusted high-rate source of qubits encoded in photons. Thus, we consider an MDI randomness expansion protocol~\cite{chaturvedi2015measurement,nie2016experimental,bischof2017measurement,guo2019experimental}.

To illustrate the potential of the proposed solution, we compare it with the recent leading result of MDI protocol secure against a quantum adversary~\cite{wang2023provably}. The implementation presented in that work has an event rate of $2.5$~MHz. The protocol was executed with $10^{10}$ rounds, requiring about $4000 s$. The net and gross randomness generation rates were $0.00199$ and $0.00455$ bits per round, respectively, yielding $4.98$~Kbps net efficiency, and the ratio between consumption and generation was $0.44$. This means that the randomness generated by the experiment needed almost half its amount to be provided from an independent source, posing a serious practical problem. The earlier experiment reported a rate of $5.7$~Kbps with an event rate of $25$~MHz, but the analysis was secure only against classical adversaries~\cite{nie2016experimental}.

To show the advantage of the approach with POVM, we assume that the test rounds requiring entanglement can be executed with a rate of $1000$ events per second, which was achieved e.g. in recent experiments with high fidelity entanglement~\cite{seguinard2023experimental}. We consider a case with the total duration of the testing and generation rounds to be $4000s$ and event rate of $2.5$~MHz, i.e. both are the same as in~\cite{wang2023provably}; thus $\gamma = 0.0004$. In contrast to~\cite{wang2023provably}, for the generation round, there is no need to create entangled pairs of photons, and thus the photon creation rate could be significantly higher. The net randomness generation rate for these parameters is around $1.32$~Mbps for $\beta = 2 \times 10^{-8}$ in~\eqref{eqs:EAT_functions}. The randomness generation requires input randomness per event $4 \gamma + H_2(\gamma) \approx 0.006692$, where $H_2(x) = -x \log_2(x) - (1-x) \log_2(1-x)$ is the binary Shannon entropy. The ratio between consumption and generation is about $0.0127$, giving a significant improvement over~\cite{wang2023provably}.

The considered event rate of $2.5$~MHz is much lower than the $312.5$~MHz round rate presented in the very recent work and giving $23$~MHz of bit rate~\cite{nie2024measurement}. With these estimations, one can expect that employing POVMs in that setup would provide about $5$ times better bit rate. We compare the parameters of selected MDI protocols in Tab.~\ref{tab:mdi}.

\begin{table}[htbp]
	\begin{tabular}{|l|l|l|l|l|}
		\hline
		paper               & event rate   & \thead{raw bit \\ per event} & bit rate     & \thead{bit rate over \\ event rate} \\ \hline
		\cite{nie2016experimental} & 25 MHz       & 0.00023           & 5.7 Kbps     & 0.000228            \\ \hline
		%\cite{guo2019experimental} & not reported & 1.106             & not reported & not reported        \\ \hline
		\cite{wang2023provably}    & 2.5 MHz      & 0.00455           & 4.98 Kbps    & 0.002               \\ \hline
		\cite{nie2024measurement} & 312.5 MHz    & 0.0737            & 23 Mbps      & 0.0736              \\ \hline
		\makecell{this work \\ (hypothe- \\ -tical)}        & 2.5 MHz      & 1.14              & 1.32 Mbps     & 0.52743 \\ \hline
	\end{tabular}
	\caption{\label{tab:mdi} Comparison of selected Measurement-DI QRNG protocols parameters with a hypothetical implementation illustrating the results of this work.}
\end{table}

\section{Conclusions}

Our approach offers advantages over previous attempts for quantum randomness certification. Our analysis of data from~\cite{smania2020experimental} shows that, using generalized measurements on a single-qubit system, $1.21$~bits of min-entropy can be obtained. It is more efficient since it requires only a single measurement on a qubit system, whereas previous protocols required multiple measurements on qubit systems to get more than 1 bit of randomness. Our results have important implications for quantum metrology, cryptography, and other areas of quantum information theory. They can be used in conjunction with the EAT to improve the certification of quantum sources of randomness and enhance the security of quantum cryptographic protocols. It would be interesting to investigate how our approach can be generalized to various types of quantum systems, including qutrits and higher-dimensional systems.

\section*{Acknowledgements}
This work was supported by the Knut and Alice Wallenberg Foundation through the Wallenberg Centre for Quantum Technology (WACQT), and the Swedish Research Council (VR).

%\bibliography{povm_randomness_refs}
\input{povm_randomness_arXiv.bbl}

\end{document}